\documentclass[twocolumn,showpacs,preprintnumbers,amsmath,amssymb,superscriptaddress,prl]{revtex4-1}

\usepackage{graphicx}
\usepackage{dcolumn}
\usepackage{bm}
\usepackage{graphics}
\usepackage{subfigure}
\usepackage{graphicx}
\usepackage{epsfig}
\usepackage{epstopdf}

\begin{document}

\title{Coarsening dynamics of binary Bose condensates}

\author{Johannes Hofmann}

\email{hofmann@umd.edu}

\author{Stefan S. Natu}

\email{snatu@umd.edu}

\author{S. Das Sarma}

\affiliation{Condensed Matter Theory Center and Joint Quantum Institute, Department of Physics, University of Maryland, College Park, Maryland 20742-4111 USA}

\date{\today}

\begin{abstract}
We study the dynamics of domain formation and coarsening in a binary Bose-Einstein condensate that is quenched across a miscible-immiscible phase transition. The late-time evolution of the system is universal and governed by scaling laws for the correlation functions. We numerically determine the scaling forms and extract the critical exponents that describe the growth rate of domain size and autocorrelations. Our data is consistent with inviscid hydrodynamic domain growth, which is governed by a universal dynamical critical exponent of $1/z = 0.68(2)$. In addition, we analyze the effect of domain wall configurations which introduce a nonanalytic term in the short-distance structure of the pair correlation function, leading to a high-momentum ``Porod''-tail in the static structure factor, which can be measured experimentally. 
\end{abstract}

\maketitle

The thermodynamic ground state of a system that consists of multiple species is not always spatially homogeneous. Indeed, as the thermodynamic state variables or the interspecies couplings are tuned, often a transition between a miscible and an immiscible ground state takes place \cite{stenger98}. A system that is quenched across such a transition does not phase-separate instantly but exhibits  highly nontrivial dynamics which generally proceed in two stages \cite{ao2000, zurek85, kibble76, bray94}: first, domains of one species nucleate over a short time-scale. In the second stage, these domains merge and coarsen until in the infinite-time limit, only one large domain of each species remains. In certain cases, the dynamics in the latter stage can be universal in that they \emph{do not depend on the microscopic details of the system} and are only constrained by symmetries and conservation laws~\cite{bray94,hohenberg77}. The time evolution is then self-similar, i.e., the time dependence of any ensemble averaged-quantity is captured by a simple rescaling of units by a characteristic length scale $L(t)$ (for example, the average domain size). In classical theories of phase ordering kinetics, this scale diverges with time according to a characteristic power law $L(t) \sim t^{1/z}$. The phase ordering dynamics of different systems can thus be separated into distinct dynamical universality classes that are characterized by the \emph{dynamical critical exponent} $z$. The concept of scaling applied to the late-time coarsening is truly universal and has found its application to a wide variety of distinct problems in physics: originally developed to describe the growth of metallic grain boundaries~\cite{mullins56} and the spinoidal decomposition of a binary alloys below a critical phase-coexistence temperature~\cite{cahn61}, scaling concepts are now used to describe to formation of galaxies, the domain growth of liquid membranes~\cite{stanich13}, or even sociopysics~\cite{castellano09}. Here, we extend the classical paradigm of phase ordering kinetics to quantum systems by presenting an example of a quantum system that exhibits classical late-time scaling: we calculate the dynamical critical exponent for domain coarsening in a  binary superfluid in two dimensions, finding a dynamical scaling exponent that is consistent with inviscid hydrodynamic domain growth. Understanding the late time dynamics of isolated quantum systems undergoing unitary (energy conserving) time evolution is an active topic of research, our work shows that established paradigms from classical coarsening can further our understanding of this fundamental problem.

Recent experiments in spinor Bose-Einstein condensates (BECs) \cite{hall98, stenger98, sadler06, guzman11, de13, papp08} (see \cite{stamper13} for a review) have made it possible to answer long standing questions about the properties of multicomponent superfluids~\cite{graf67}. These ultra-cold gases are unique quantum fluids as they are largely isolated from the environment and their quantum dynamics can be probed accurately over long times both \textit{in-situ} \cite{sadler06, parker13} and in time-of-flight \cite{hall98, stenger98, de13, papp08}. Furthermore, the unprecedented control offered by quantum gas experiments makes them an ideal testbed to study the nonequilibrium physics of superfluids following a parameter quench \cite{karl13, karl132, nowak12, nowak11, schole12}. Indeed, the short-time dynamics of domain formation following a quench are well understood both theoretically and experimentally \cite{hall98, stenger98, sadler06, stamper13, kasamatsu04, lamacraft07, ronen08, barnett11}. However, questions remain about the universal, long time dynamics and the mechanisms governing domain growth \cite{mukerjee07, lamacraft07, kudo13, damle96, nowak12, nowak11, schole12, karl13, karl132}. Surprisingly, even in the seemingly simple binary Bose system we consider, the power law governing \emph{long-time} domain growth at zero temperature was hitherto unknown. Here, we theoretically establish that the dynamical critical exponent to be $1/z = 0.68 (2)$.

\begin{figure*}
\subfigure[$\, \hat{t}=300$]{\raisebox{0.35cm}{\scalebox{0.27}{\includegraphics{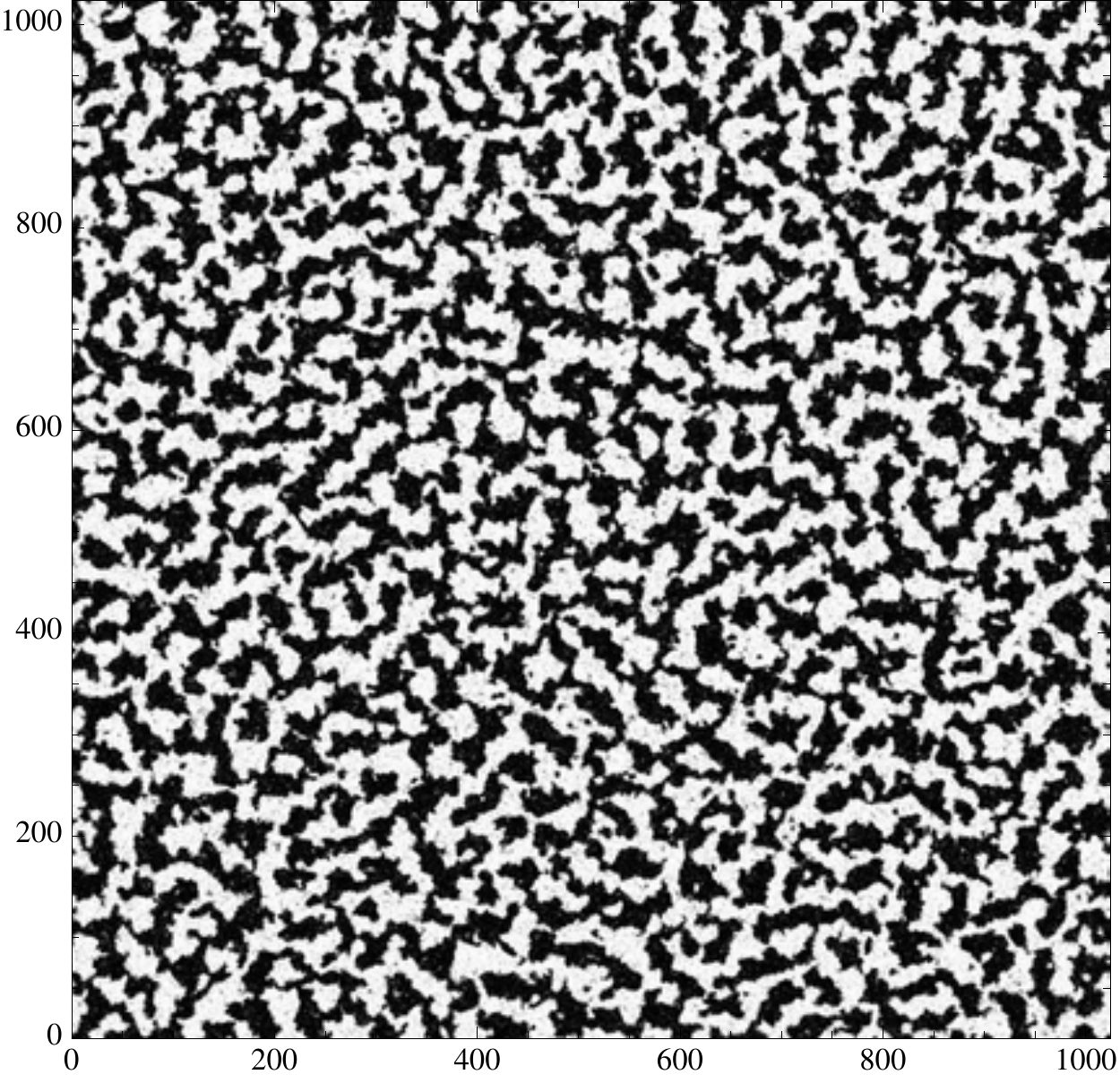}}}}
\subfigure[$\, \hat{t}=1000$]{\raisebox{0.35cm}{\scalebox{0.27}{\includegraphics{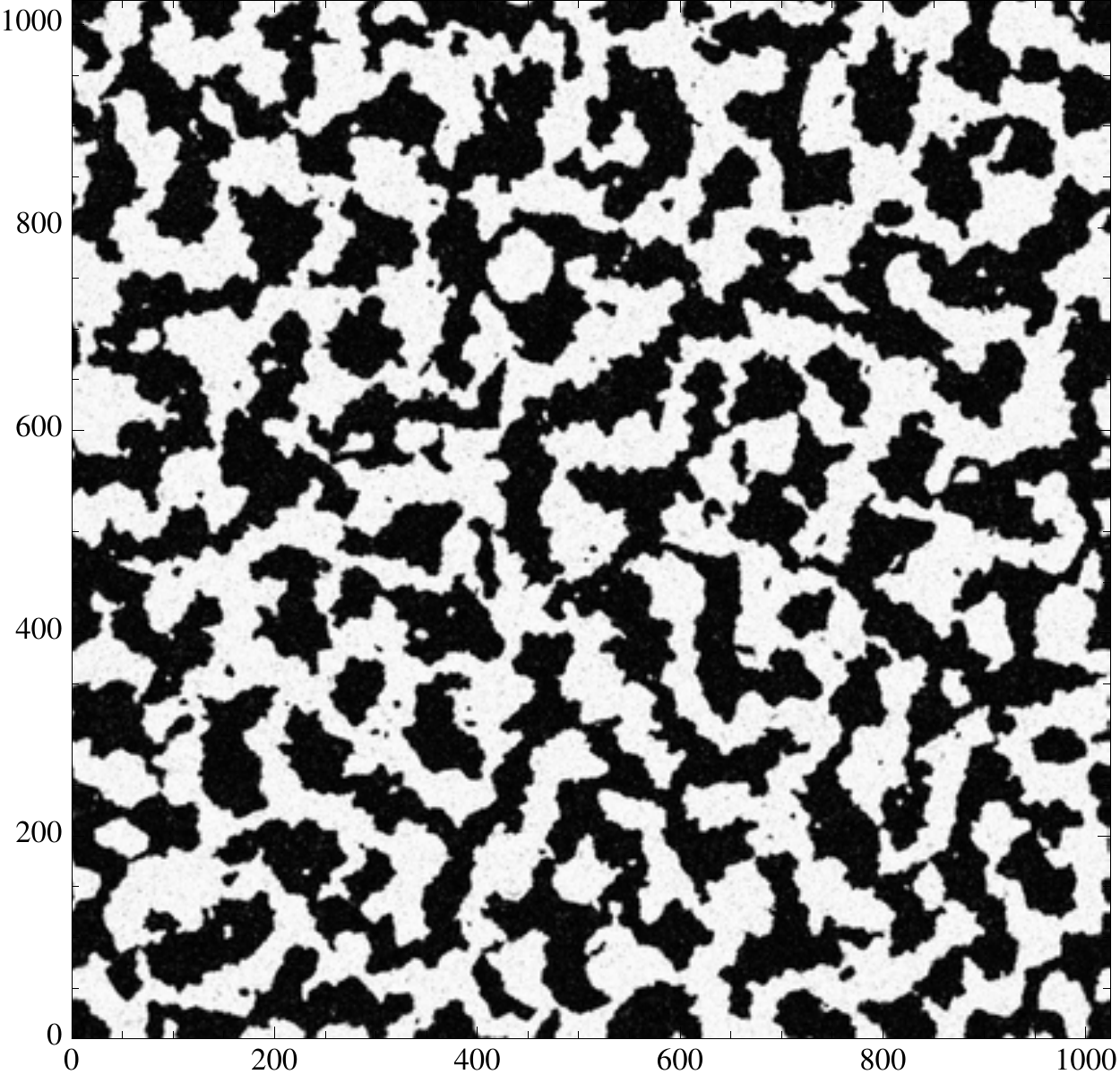}}}}
\subfigure[$\, \hat{t}=2000$]{\raisebox{0.35cm}{\scalebox{0.27}{\includegraphics{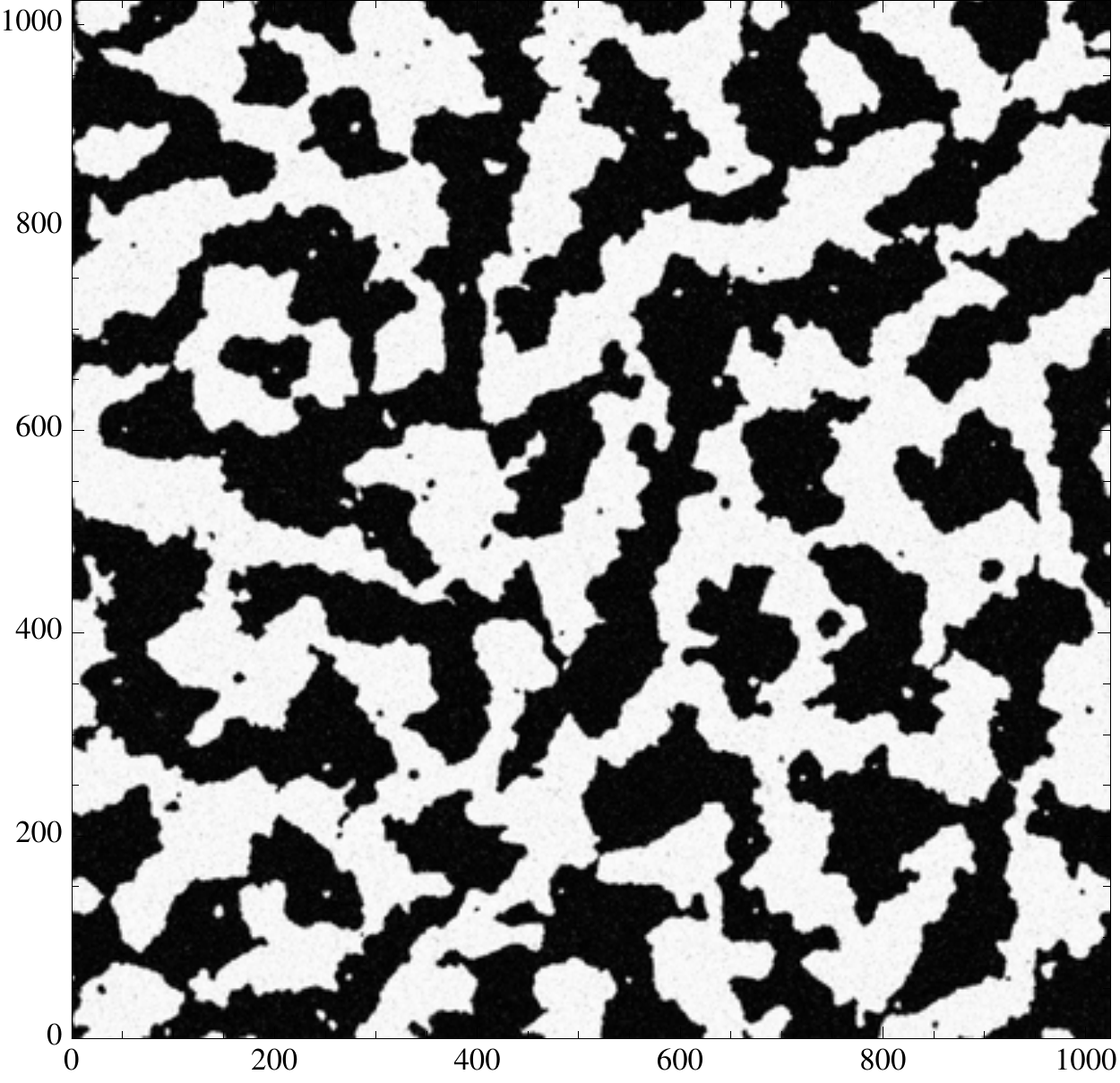}}}}
\subfigure[$\, \hat{t}=5000$]{\raisebox{0.35cm}{\scalebox{0.27}{\includegraphics{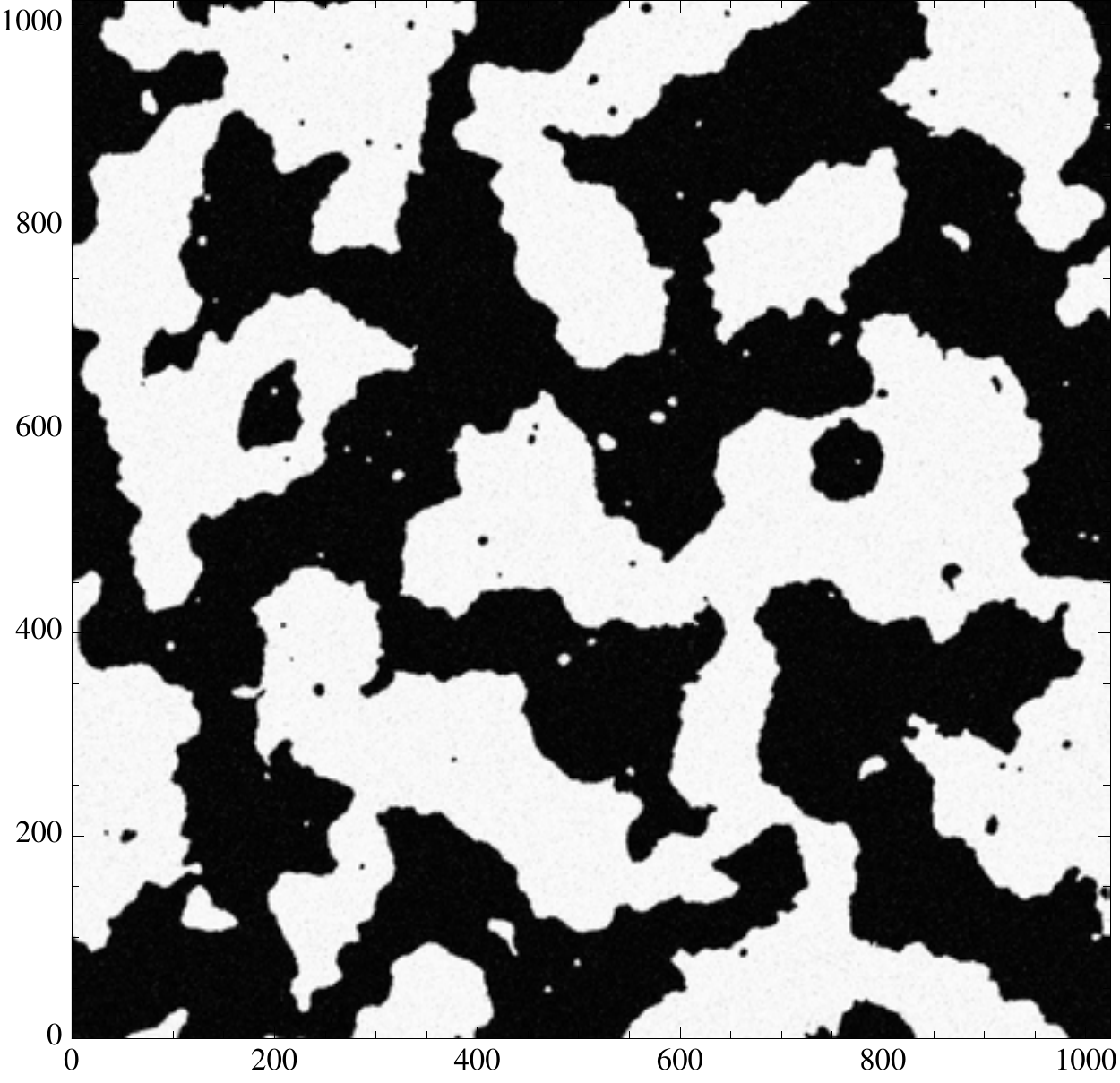}}}}
\subfigure[]{\scalebox{0.30}{\includegraphics{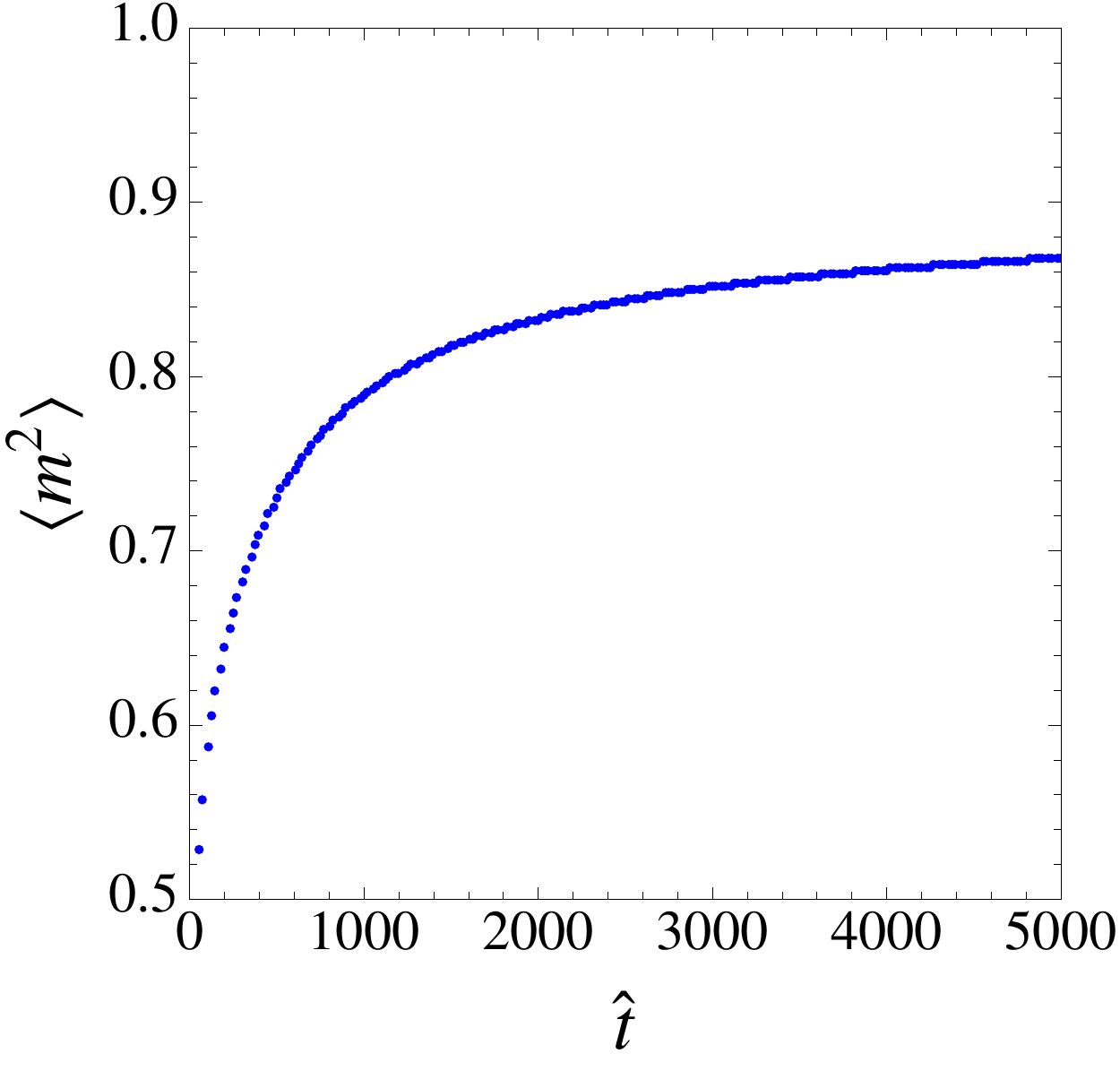}}}
\caption{
(a-d) Snapshot of the time evolution of a system on a $1024$ square lattice at times $\hat{t}=300,1000,2000$, and $5000$. Domains of positive (negative) magnetization are shown in black (white). After the system is quenched across the instability, domains begin to form which then undergo a self-similar coarsening evolution that is governed by universal scaling laws. (e) Average quadratic magnetization $\langle m^2 \rangle$ as a function of time. The magnetization of each domain grows with time.
}
\label{fig:evolution}
\end{figure*}

We describe the binary BEC at zero temperature by two classical fields $\psi_1$ and $\psi_2$, and take into account the interaction between the particles on a mean field level. In typical ultra-cold gas experiments, these fields correspond to two hyperfine states of an atom such as $^{23}$Na or $^{87}$Rb (\cite{hall98, stenger98, de13}), or two different atomic species ($^{87}$Rb and $^{85}$Rb) as in the experiment of Papp \textit{et al.} \cite{papp08}. This yields the Gross-Pitaevskii (GP) energy functional:

\begin{align}
H &= \int d^2r \, \Bigl(\sum_{i=1,2} \Big\{-\psi^\dagger_i \frac{\nabla^2}{2m_i} \psi_i - \mu_i |\psi_i|^2+ \frac{g_{ii}}{2} |\psi_i|^4 \Bigr\} \nonumber \\
& \qquad + g_{12} |\psi_1|^2 |\psi_2|^2 \Bigr) , \label{eq:hamiltonian}
\end{align}

where $m_i$ is the particle mass, $\mu_i$ the chemical potential, $g_{ii}$ is the interaction between like particles of type $i$, and $g_{12}$ describes the scattering of atoms of spin $1$ and $2$. In the absence of long-range dipolar interactions \cite{sadler06}, the interaction Hamiltonian does not contain spin-flip terms and conserves the total density of each species. Note that the GP equation is strictly applicable only at zero temperature and does not account for quantum fluctuations. More complex models describing the time-evolution of the superfluid mixture, such as the ``Model F'' in the Hohenberg-Halperin classification~\cite{hohenberg77}, reduce to the GP equation in the low-temperature limit~\cite{mukerjee07}. Here, we model the GP equation directly and discuss the limitations of our approach later.

In typical spinor BECs such as the hyperfine states of $^{87}$Rb or $^{23}$Na, the scattering length is nearly identical in all channels~\cite{kempen02}, and we choose $g_{11}=g_{22}=g>0$ in the following. If the intra-species interaction dominates, $g_{12}< \sqrt{g_{11}g_{22}}$, the two condensates can coexist. However, if $g_{12}>\sqrt{g_{11}g_{22}}$, the ground state is no longer homogeneous and the system phase-separates~\cite{colson78,law97,hall98,stamper99}. We investigate how the system evolves when suddenly quenched from the miscible phase with $g_{12} < g$ to the immiscible phase with $g_{12}>g$. Experimentally, this can be done either using a Feshbach resonance in systems such as $^{85}$Rb-$^{87}$Rb mixtures \cite{chengreview2010}, or by preparing a miscible initial state in an otherwise immisicible mixture (such as the hyperfine states of $^{87}$Rb or $^{23}$Na \cite{stenger98, hall98}) using a transverse magnetic field and observing the subsequent dynamics \cite{de13}. We do not expect that our conclusions will be modified if $g_{11} \ne g_{22}$, as the precise mechanism for coarsening does not depend on the specific choice of interaction parameters. 

We consider the time-evolution of the polarization $m({\bf r}) = (n_1({\bf r}) - n_2({\bf r}))/(n_1({\bf r}) + n_2({\bf r}))$ as an order parameter. The spin texture of a spinor gas, and thus the magnetization, can be measured directly using spin-sensitive phase contrast imaging~\cite{guzman11,de13}. At early times, domains of opposite spin form due to a spin-wave instability~\cite{kasamatsu04}. Taking into account only the most unstable mode, the initial domain size is of order $L_0\approx \xi_s$, where we define the spin healing length $\xi_s=\sqrt{1/2 m n (g_{12}-g)}$, and $n=|\psi_1|^2 + |\psi_2|^2$ is the total density. When the domain size becomes much larger than the spin healing length, the dynamics are universal. Expressed in units of the characteristic length scale $L(t)$, all correlation functions of the order parameter $m$ have no explicit time dependence, and collapse to a single, universal scaling function. The pair correlation function, which describes the correlation of the magnetization at two points separated by a distance $r$, can be written as follows:
\begin{align}
g(r,t) &= \frac{1}{V} \int d^2R \, \langle m(R) m(R+r) \rangle = f(r L^{-1}(t)) .\label{eq:paircorr}
\end{align}
The bracket $\langle \ldots \rangle$ denotes an ensemble-average. The correlation function is normalized such that $g(0,t) = 1$. Similarly, the static structure factor assumes the scaling form
\begin{align}
S(q) &= \int d^2r \, e^{i {\bf q} \cdot {\bf r}} g(r,t) = L^2 \hat{f}(q L(t)) .
\end{align}
It should be emphasized that the scaling is a \textit{conjecture} which must be proven on a case-to-case basis \cite{bray94}.

The equations of motion corresponding to the Hamiltonian~\eqref{eq:hamiltonian} are the well-known GP equations: 
\begin{align}
i \partial_t \psi_i &=  \left\{- \frac{\nabla^2}{2m_1} + g |\psi_i|^2 + g_{12} |\psi_j|^2 \right\}  \psi_i  \label{eq:gp}
\end{align}
where $i, j = \{1, 2\}$ and $j \neq i$.
In the following, we consider two species of equal mass $m_i = m$ and equal chemical potential $\mu_i=\mu$. We simulate the time evolution~\eqref{eq:gp} on a square lattice of dimension $l = 1024$ and spacing $d$ using a split-step spectral method as introduced in Ref.~\cite{bao03}. Although we only present results for size $l = 1024$, we have performed numerical simulations over a wide range of system sizes $l = 64, 128, 512$, finding the same dynamical critical exponent in each case. We initialize the system in the ground state $\psi_i = \sqrt{\mu/(g+g_{12})}$ and add a small Gaussian noise to seed the instability after a sudden quench. The results are averaged over several noise realizations. The chemical potential is chosen such that the initial domain size ($L_0 \sim \xi_{s}$) is much larger than the lattice spacing but is still much smaller than the system size.

Figure~\ref{fig:evolution} shows the domain structure at various times $\hat{t} = md^2t=300,1000,2000$, and $5000$ for a quench with final couplings $g=1$ and $g_{12}=1.1$. In typical experiments with $^{87}$Rb or $^{23}$Na, the coupling strengths are almost identical with $(g_{12}-g)/g \approx 10^{-3}$, however these can be tuned using a Feshbach resonance~\cite{papp08, weber02, vuleti99, pollack09, papp08}. The choice of couplings determines the initial domain size after the quench and the onset of the universal late-time scaling regime, but does not affect the scaling behavior itself. Regions with positive (negative) polarization are shown in black (white). Shortly after the quench, distinct domains form (Fig.~\ref{fig:evolution}(a)) the size of which grows with time (Figs.~\ref{fig:evolution}(b)-(c)). In the following, we compute the universal scaling exponent that governs this time evolution. Note that even at late times, the domains are not fully polarized but contain small patches of opposite polarization, which correspond to vortex or skyrmion defects~\cite{karl13}.  For non-dissipative evolution, these defects are long-lived \cite{karl13}, but they decay if dissipative effects are included into the GP equations~\cite{kudo13, kasamatsu04, barnett11}.

\begin{figure}
\subfigure{\scalebox{0.51}{\includegraphics{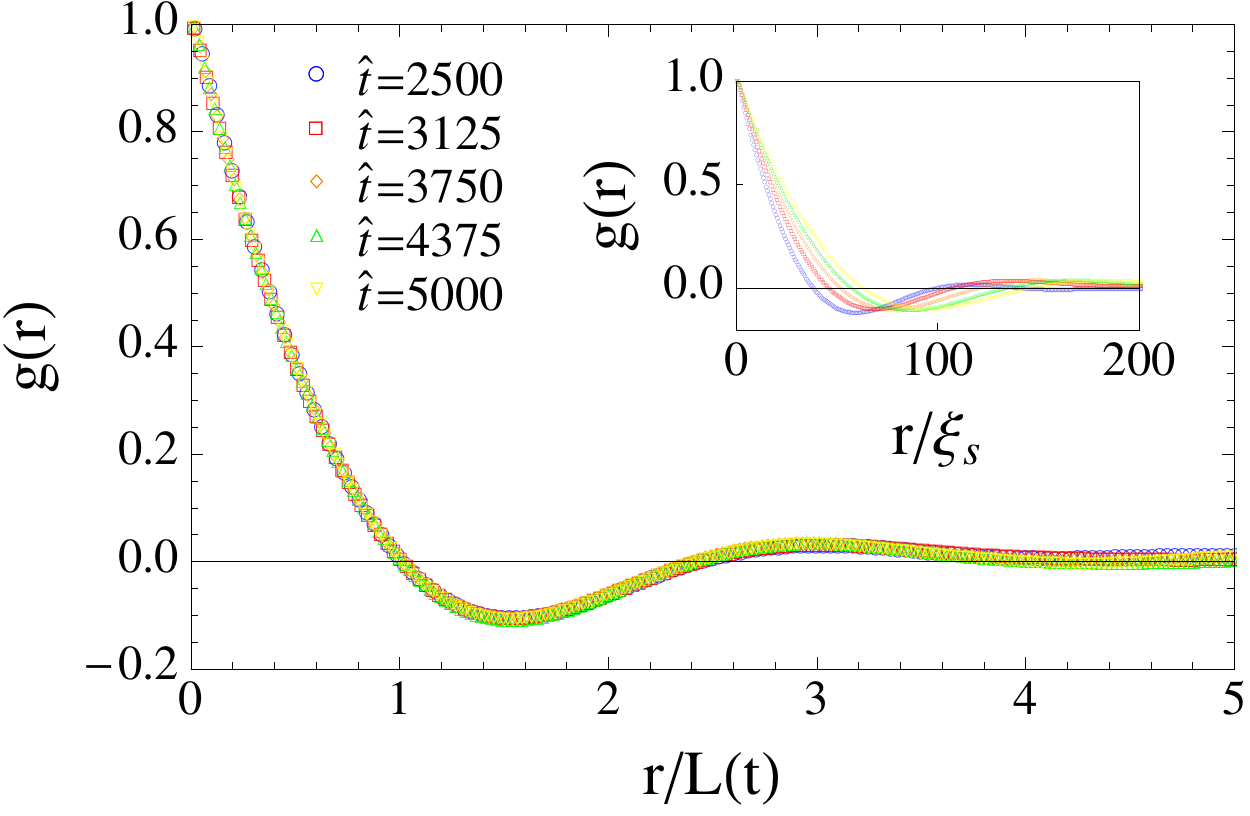}}\label{fig:paircorr}}
\, \subfigure{\scalebox{0.51}{\includegraphics{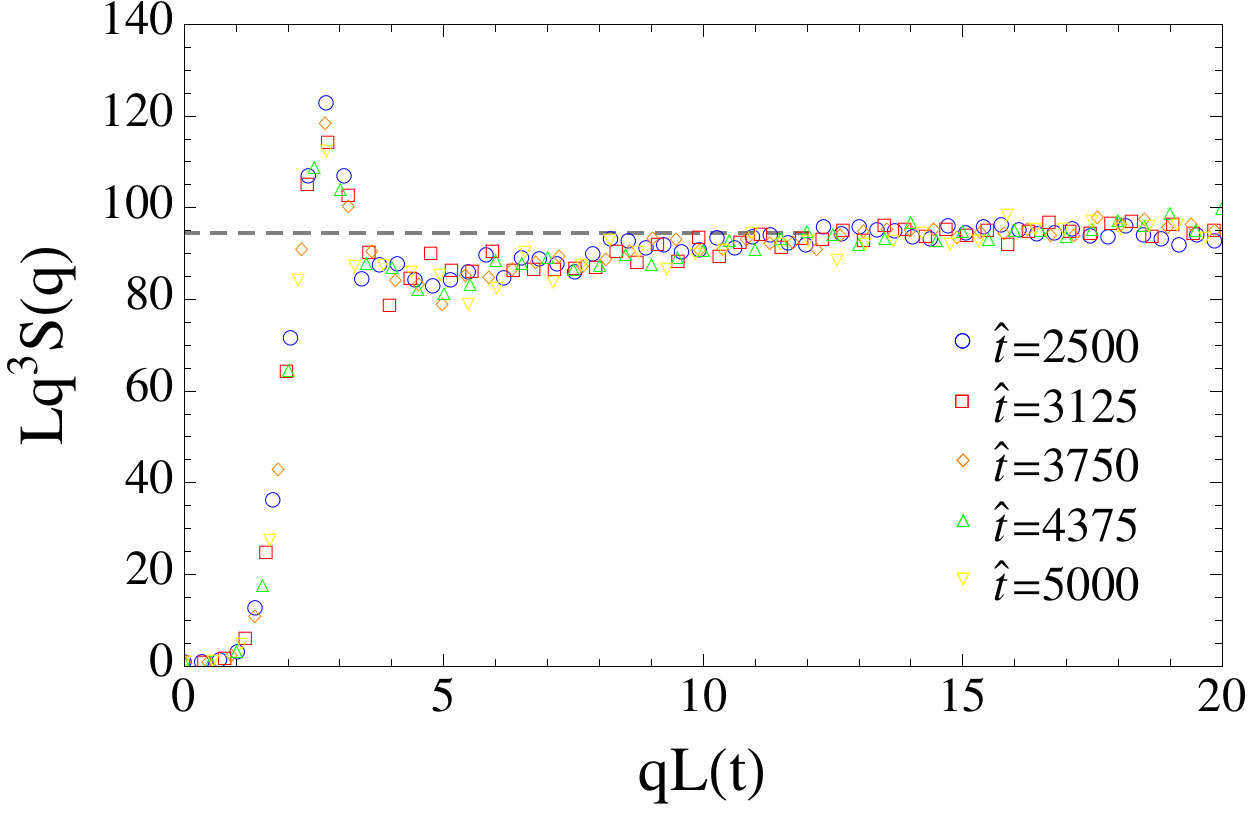}}\label{fig:structure}}
\caption{
(Color Online) (Top) Scaling function of the pair correlations function at different times. Inset: Pair correlation function. (Bottom) Universal scaling function of the structure factor at different times $\hat{t}=300,1000,1500,2000$, and $5000$. The structure factor collapses to a single function when expressed in units of the domain size. The presence of domain walls induces a high-momentum tail $S(q) \sim q^{-3}$.
}
\end{figure}

In Figure~\ref{fig:paircorr}, we show the scaling function $f$ of the pair correlation function ($g(r,t)$) at various times. The inset shows the unrescaled pair correlation function. All results are averaged over $25$ initial conditions. For small separation $r$, the correlation is positive, indicating that two nearby points have predominantly the same polarization. This changes at larger distances, where two points become anticorrelated. We take the position of the first zero in the pair correlation function as a measure for the average domain size $L(t)$. For sufficiently large domains, the results obtained from this method are consistent with those obtained by extracting the domain size directly from the simulations (Fig.~\ref{fig:evolution}). Higher oscillations are visible but very weak, and the correlation between two points is lost at large distances. As can be seen from Fig.~\ref{fig:paircorr}, when expressed in terms of the rescaled coordinate $r/L(t)$, the correlation functions at different times collapse onto a single scaling function. Indeed, the time evolution is consistent with the scaling hypothesis.

The scaling behavior is also visible in the structure factor $S(q)$ shown in Fig.~\ref{fig:structure}. Moreover, we see that the structure factor decays as a power law $\sim 1/q^3$ at large momentum. This feature is a consequence of the presence of domain walls, which determine the short-distance structure of the pair correlations. It can be understood as follows: two points with separation $r$ are positively correlated if they are in domains of the same type and negatively otherwise. At short distances, this correlation is determined by the probability to cross a single domain wall, which is proportional to $r/L(t)$~\cite{bray94}. This gives rise to a nonanalytic linear term in the pair correlation function $g(r) = 1 + f'(0) \frac{r}{L(t)} + \textit{O}(r^2)$. The linear slope is clearly present in our numerical data, Fig.~\ref{fig:paircorr}. This implies a universal high-momentum ``Porod'' tail \cite{bray94} in the structure factor $S(q) = \frac{C}{L q^3}  + \textit{O}(1/q^4)$. Crucially, this high-momentum tail is not present in the initial stage of the coarsening, where the structure factor is Gaussian. The structure factor changes from its initial Gaussian shape to the form in Fig.~\ref{fig:structure} only after the domains have formed and the system has entered the second coarsening stage. At late times when the density of domain walls decreases, the magnitude of the tail vanishes as $1/L(t)$.
 
\begin{figure}[b]
\scalebox{0.6}{\includegraphics{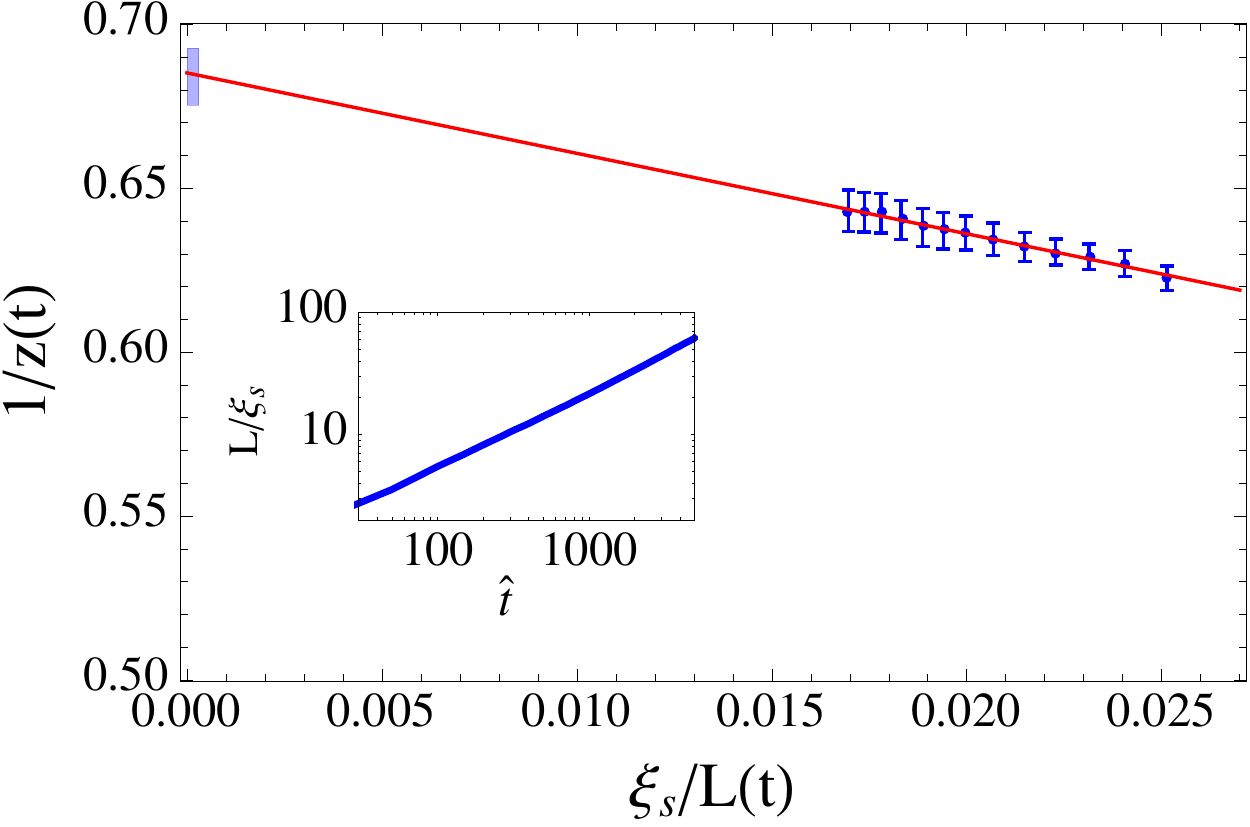}}\caption{
(color online) time-dependent dynamical scaling exponent $1/z(t)$. The linear fit to the data points (red line, see text) gives a scaling exponent $1/z(\infty) = 0.68(2)$. Inset: Log-Log plot of domain size as a function of time.}
\label{fig:zoft}
\end{figure}

We proceed to extract the scaling exponent from the measured domain sizes. The scaling argument is strictly valid only in the limit of infinitely large domain sizes, and the exponent has a finite size scaling correction. Huse argued that for a conserved order parameter, this correction is of order $1/L(t)$~\cite{huse86}. We determine the time-dependent scaling exponent $z(t)$ by taking the logarithmic differential quotient of the domain size at different times $t'>t$: $1/z(t) = \log (L'/L)/\log (t'/t)$. As is apparent from the result in Fig.~\ref{fig:zoft}, the scaling exponent displays a weak drift towards smaller values at later times. Extrapolating to the limit of infinite domain size (red line in Fig.~\ref{fig:zoft}), we obtain $1/z(\infty) = 0.68(2)$. The same dynamical exponent is found in classical binary fluids where $1/z = 2/3$~\cite{furukawa85}. 

\begin{figure}
\scalebox{0.6}{\includegraphics{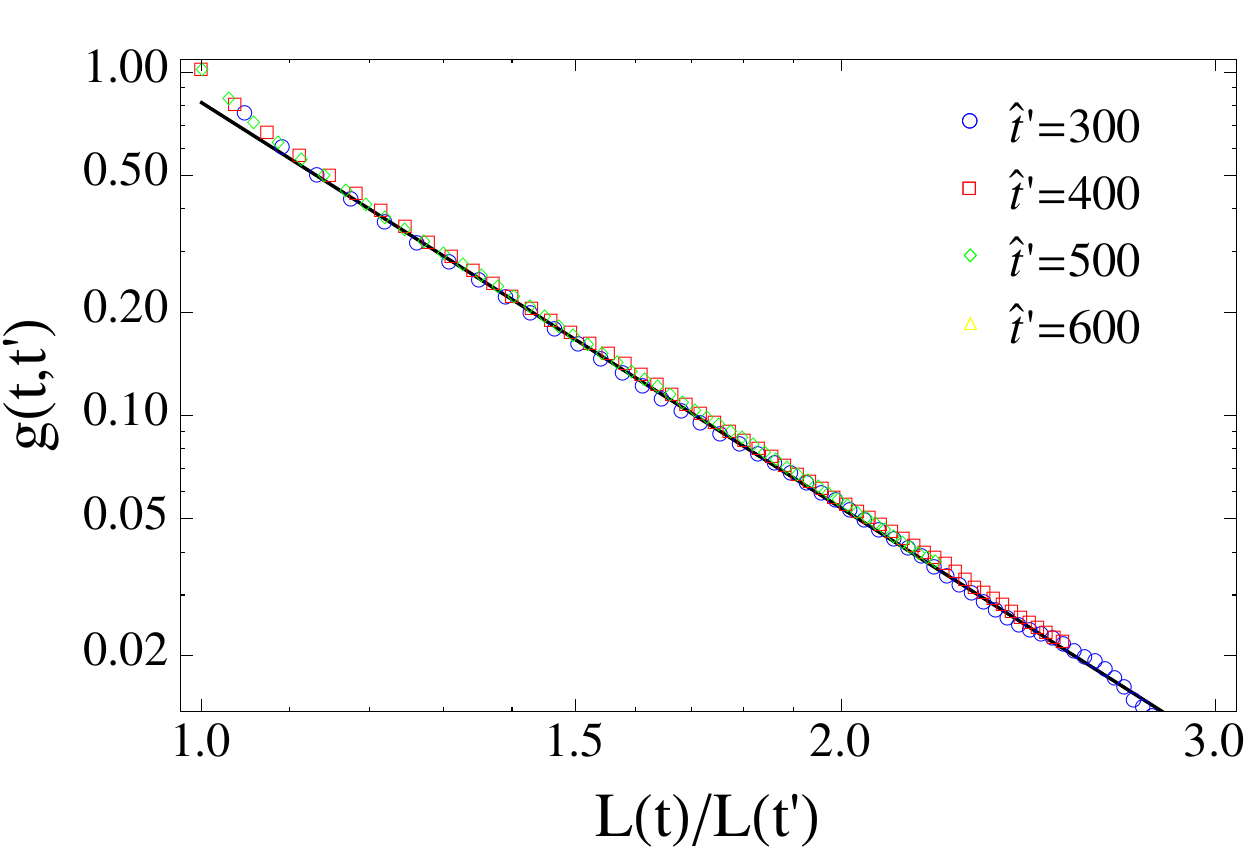}}
\caption{
(color online) Autocorrelation function $g(t;t')$ as a function of $L(t)/L(t')$ for different values of $\hat{t}'=300,400,500$, and $600$ (left to right). A linear fit (red line) gives the scaling exponent $\lambda=3.90(2)$.
}
\label{fig:autocorr}
\end{figure}

The numerical value of the scaling exponent $z$ indicates that the dominant mechanism driving domain growth is inertial hydrodynamic transport of superfluid from low-density to high-density regions \cite{bray94}. Heuristically, this can be understood as follows: in a binary fluid, the change in a domain wall's velocity $d{\bf v}/dt$ is equal to the gradient of the pressure $-\nabla P$, where the pressure $P$ is proportional to the energy density of the domain walls. The inertial term scales as $d{\bf v}/dt \sim L/t^2$, and the pressure gradient scales as $- \nabla P \sim \sigma/L^2$, where $\sigma$ is the surface tension~\cite{bert08}. Equating inertial and gradient terms gives the dynamical exponent $1/z=2/3$ \cite{bray94}. This exponent has not been measured in classical binary liquids primarily because coarsening dynamics is complicated by advective and viscous terms which give rise to other scaling exponents, and can even lead to a  breakdown of scale invariance \cite{siggia79, wagner98}. However binary BECs may be the ideal testbed for observing this type of scaling behavior.

Our previous analysis was restricted to the equal-time correlation function. It turns out that a single dynamical exponent is not sufficient to describe the scaling behavior of correlators at different times \cite{bray94}. Consider the dynamical pair correlation function, which in the scaling limit depends on two length scales:
\begin{align}
g(r,t;r',t') &= \frac{\langle m(r,t) m(r',t') \rangle}{\langle m^2(r,t) \rangle^{1/2} \langle m^2(r',t') \rangle^{1/2}} \nonumber \\
&= f\left(\frac{r}{L(t)}, \frac{r'}{L(t')}\right) .
\end{align}
In the limit of $t\gg t'$, the dependence on one length scale separates, and the correlation function becomes a homogeneous function of the ratio $L(t)/L'(t)$:
\begin{align}
g(r,t;r',t') &= \left(\frac{L(t)}{L(t')}\right)^{-\lambda} \hat{f}(r/L(t)) .
\end{align}
In general, the exponent $\lambda$ differs from $z$. We show the behavior of the zero-range value of the pair correlation function $g(0,t;0,t')$ as a function of time in Fig.~\ref{fig:autocorr}. We assume that the correction arising from the finite domain size is small and perform a direct fit to the data over the whole measurement interval. This procedure gives a dynamical exponent of $\lambda = 3.90(2)$.

We now discuss the significance of our results to non-equilibrium experiments in binary Bose condensates. In the experiments to date, the spin healing length is of order $\zeta \sim 10\mu$m \cite{sadler06, stenger98, hall98}, which corresponds to a time of $t \sim 1$s beyond which scaling should be observed. On these timescales, the dynamics of the physical system is complicated by particle losses and heating in the trap in present-day experiments~\cite{de13}, which however should not preclude the observation of universal late-time scaling in future, higher-precision experiments. Another promising approach for realizing an effective strongly interacting spin-$1/2$ system is to use lattice modulation \cite{parker13}. While stronger interactions imply smaller spin healing lengths, inelastic collisions, three-body losses and heating may complicate the observation of universal scaling. 

Finally, even at temperatures $T \ll \mu$, where the GP approach is valid, on long timescales of order $t_{\text{coll}} \sim (\hbar a n/ml_{z} \sqrt{(na^{3}/l_{z})})^{-1} \gg m \xi_s^{-2}$, (where $l_{z}$ is the axial confinement length), inelastic scattering processes will start to become important, in which case the system should be modeled using a quantum kinetic equation rather than the GP ansatz. A proper modeling of inelastic effects requires more sophisticated methods such as the c-field techniques currently being developed by several groups \cite{berloff02,blakie08}. While coarsening may still be observed on times $t < t_{\text{coll}}$, it remains an exciting future problem to investigate the transition between the Gross-Pitaevskii and kinetic limit.
 
In conclusion, we provide numerical evidence that the coarsening dynamics of a binary Bose-Einstein condensate quenched across a miscible-immiscible phase boundary obeys universal scaling laws. Equal-time correlation functions depend implicitly on time through a single characteristic length scale which grows according to a power law with time with an exponent $1/z = 0.68(2)$, consistent with inertial hydrodynamic growth. Our findings can be verified experimentally either via direct imaging of magnetic domain growth \cite{hall98, stenger98, sadler06, de13, parker13}, or by measuring the structure factor \cite{miyake11}.

\textit{Acknowledgements.---} This work is supported by JQI-NSF-PFC, AFOSR-MURI, and ARO-MURI. 

\bibliography{bib}

\begin{thebibliography}{10}%
\makeatletter
\providecommand \@ifxundefined [1]{%
 \ifx #1\undefined \expandafter \@firstoftwo
 \else \expandafter \@secondoftwo
\fi
}%
\providecommand \@ifnum [1]{%
 \ifnum #1\expandafter \@firstoftwo
 \else \expandafter \@secondoftwo
\fi
}%
\providecommand \enquote [1]{``#1''}%
\providecommand \bibnamefont  [1]{#1}%
\providecommand \bibfnamefont [1]{#1}%
\providecommand \citenamefont [1]{#1}%
\providecommand\href[0]{\@sanitize\@href}%
\providecommand\@href[1]{\endgroup\@@startlink{#1}\endgroup\@@href}%
\providecommand\@@href[1]{#1\@@endlink}%
\providecommand \@sanitize [0]{\begingroup\catcode`\&12\catcode`\#12\relax}%
\@ifxundefined \pdfoutput {\@firstoftwo}{%
 \@ifnum{\z@=\pdfoutput}{\@firstoftwo}{\@secondoftwo}%
}{%
 \providecommand\@@startlink[1]{\leavevmode\special{html:<a href="#1">}}%
 \providecommand\@@endlink[0]{\special{html:</a>}}%
}{%
 \providecommand\@@startlink[1]{%
  \leavevmode
  \pdfstartlink
   attr{/Border[0 0 1 ]/H/I/C[0 1 1]}%
   user{/Subtype/Link/A<</Type/Action/S/URI/URI(#1)>>}%
  \relax
 }%
 \providecommand\@@endlink[0]{\pdfendlink}%
}%
\providecommand \url  [0]{\begingroup\@sanitize \@url }%
\providecommand \@url [1]{\endgroup\@href {#1}{\urlprefix}}%
\providecommand \urlprefix [0]{URL }%
\providecommand \Eprint[0]{\href }%
\@ifxundefined \urlstyle {%
  \providecommand \doi [1]{doi:\discretionary{}{}{}#1}%
}{%
  \providecommand \doi [0]{doi:\discretionary{}{}{}\begingroup
  \urlstyle{rm}\Url }%
}%
\providecommand \doibase [0]{http://dx.doi.org/}%
\providecommand \Doi[1]{\href{\doibase#1}}%
\providecommand \bibAnnote [3]{%
  \BibitemShut{#1}%
  \begin{quotation}\noindent
    \textsc{Key:}\ #2\\\textsc{Annotation:}\ #3%
  \end{quotation}%
}%
\providecommand \bibAnnoteFile [2]{%
  \IfFileExists{#2}{\bibAnnote {#1} {#2} {\input{#2}}}{}%
}%
\providecommand \typeout [0]{\immediate \write \m@ne }%
\providecommand \selectlanguage [0]{\@gobble}%
\providecommand \bibinfo [0]{\@secondoftwo}%
\providecommand \bibfield [0]{\@secondoftwo}%
\providecommand \translation [1]{[#1]}%
\providecommand \BibitemOpen[0]{}%
\providecommand \bibitemStop [0]{}%
\providecommand \bibitemNoStop [0]{.\EOS\space}%
\providecommand \EOS [0]{\spacefactor3000\relax}%
\providecommand \BibitemShut [1]{\csname bibitem#1\endcsname}%
\bibitem{stenger98}%
  \BibitemOpen
  \bibfield{author}{%
  \bibinfo {author} {\bibfnamefont{J.}~\bibnamefont{Stenger}}, \bibinfo
  {author} {\bibfnamefont{S.}~\bibnamefont{Inouye}}, \bibinfo {author}
  {\bibfnamefont{D.~M.}\ \bibnamefont{Stamper-Kurn}}, \bibinfo {author}
  {\bibfnamefont{H.-J.}\ \bibnamefont{Miesner}}, \bibinfo {author}
  {\bibfnamefont{A.~P.}\ \bibnamefont{Chikkatur}},\ and\ \bibinfo {author}
  {\bibfnamefont{W.}~\bibnamefont{Ketterle}},\ }%
  \bibfield{journal}{%
  \bibinfo {journal} {Nature (London)}\ }%
  \textbf{\bibinfo {volume} {396}},\ \bibinfo {pages} {345} (\bibinfo {year}
  {1998})%
  \bibAnnoteFile{NoStop}{stenger98}%
\bibitem{ao2000}%
  \BibitemOpen
  \bibfield{author}{%
  \bibinfo {author} {\bibfnamefont{P.}~\bibnamefont{Ao}}\ and\ \bibinfo
  {author} {\bibfnamefont{S.~T.}\ \bibnamefont{Chui}},\ }%
  \bibfield{journal}{%
  \bibinfo {journal} {Journal of Physics B: Atomic, Molecular and Optical
  Physics}\ }%
  \textbf{\bibinfo {volume} {33}},\ \bibinfo {pages} {535} (\bibinfo {year}
  {2000})%
  \bibAnnoteFile{NoStop}{ao2000}%
\bibitem{zurek85}%
  \BibitemOpen
  \bibfield{author}{%
  \bibinfo {author} {\bibfnamefont{W.~H.}\ \bibnamefont{Zurek}},\ }%
  \bibfield{journal}{%
  \bibinfo {journal} {Nature (London)}\ }%
  \textbf{\bibinfo {volume} {317}},\ \bibinfo {pages} {505} (\bibinfo {year}
  {1985})%
  \bibAnnoteFile{NoStop}{zurek85}%
\bibitem{kibble76}%
  \BibitemOpen
  \bibfield{author}{%
  \bibinfo {author} {\bibfnamefont{T.~W.~B.}\ \bibnamefont{Kibble}},\ }%
  \bibfield{journal}{%
  \bibinfo {journal} {Journal of Physics A: Mathematical and General}\ }%
  \textbf{\bibinfo {volume} {9}},\ \bibinfo {pages} {1387} (\bibinfo {year}
  {1976})%
  \bibAnnoteFile{NoStop}{kibble76}%
\bibitem{bray94}%
  \BibitemOpen
  \bibfield{author}{%
  \bibinfo {author} {\bibfnamefont{A.}~\bibnamefont{Bray}},\ }%
  \bibfield{journal}{%
  \bibinfo {journal} {Advances in Physics}\ }%
  \textbf{\bibinfo {volume} {43}},\ \bibinfo {pages} {357} (\bibinfo {year}
  {1994})%
  \bibAnnoteFile{NoStop}{bray94}%
\bibitem{hohenberg77}%
  \BibitemOpen
  \bibfield{author}{%
  \bibinfo {author} {\bibfnamefont{P.~C.}\ \bibnamefont{Hohenberg}}\ and\
  \bibinfo {author} {\bibfnamefont{B.~I.}\ \bibnamefont{Halperin}},\ }%
  \bibfield{journal}{%
  \bibinfo {journal} {Rev. Mod. Phys.}\ }%
  \textbf{\bibinfo {volume} {49}},\ \bibinfo {pages} {435} (\bibinfo {year}
  {1977})%
  \bibAnnoteFile{NoStop}{hohenberg77}%
\bibitem{mullins56}%
  \BibitemOpen
  \bibfield{author}{%
  \bibinfo {author} {\bibfnamefont{W.~W.}\ \bibnamefont{Mullins}},\ }%
  \bibfield{journal}{%
  \bibinfo {journal} {Journal of Applied Physics}\ }%
  \textbf{\bibinfo {volume} {27}},\ \bibinfo {pages} {900} (\bibinfo {year}
  {1956})%
  \bibAnnoteFile{NoStop}{mullins56}%
\bibitem{cahn61}%
  \BibitemOpen
  \bibfield{author}{%
  \bibinfo {author} {\bibfnamefont{J.~W.}\ \bibnamefont{Cahn}},\ }%
  \bibfield{journal}{%
  \bibinfo {journal} {Acta Metallurgica}\ }%
  \textbf{\bibinfo {volume} {9}},\ \bibinfo {pages} {795} (\bibinfo {year}
  {1961})%
  \bibAnnoteFile{NoStop}{cahn61}%
\bibitem{stanich13}%
  \BibitemOpen
  \bibfield{author}{%
  \bibinfo {author} {\bibfnamefont{C.~A.}\ \bibnamefont{Stanich}}, \bibinfo
  {author} {\bibfnamefont{A.~R.}\ \bibnamefont{Honerkamp-Smith}}, \bibinfo
  {author} {\bibfnamefont{G.~G.}\ \bibnamefont{Putzel}}, \bibinfo {author}
  {\bibfnamefont{C.~S.}\ \bibnamefont{Warth}}, \bibinfo {author}
  {\bibfnamefont{A.~K.}\ \bibnamefont{Lamprecht}}, \bibinfo {author}
  {\bibfnamefont{P.}~\bibnamefont{Mandal}}, \bibinfo {author}
  {\bibfnamefont{E.}~\bibnamefont{Mann}}, \bibinfo {author}
  {\bibfnamefont{T.-A.~D.}\ \bibnamefont{Hua}},\ and\ \bibinfo {author}
  {\bibfnamefont{S.~L.}\ \bibnamefont{Keller}},\ }%
  \bibfield{journal}{%
  \bibinfo {journal} {Biophysical Journal}\ }%
  \textbf{\bibinfo {volume} {105}},\ \bibinfo {pages} {444} (\bibinfo {year}
  {2013})%
  \bibAnnoteFile{NoStop}{stanich13}%
\bibitem{castellano09}%
  \BibitemOpen
  \bibfield{author}{%
  \bibinfo {author} {\bibfnamefont{C.}~\bibnamefont{Castellano}}, \bibinfo
  {author} {\bibfnamefont{S.}~\bibnamefont{Fortunato}},\ and\ \bibinfo {author}
  {\bibfnamefont{V.}~\bibnamefont{Loreto}},\ }%
  \bibfield{journal}{%
  \bibinfo {journal} {Rev. Mod. Phys.}\ }%
  \textbf{\bibinfo {volume} {81}},\ \bibinfo {pages} {591} (\bibinfo {month}
  {May}\ \bibinfo {year} {2009})%
  \bibAnnoteFile{NoStop}{castellano09}%
\bibitem{hall98}%
  \BibitemOpen
  \bibfield{author}{%
  \bibinfo {author} {\bibfnamefont{D.~S.}\ \bibnamefont{Hall}}, \bibinfo
  {author} {\bibfnamefont{M.~R.}\ \bibnamefont{Matthews}}, \bibinfo {author}
  {\bibfnamefont{J.~R.}\ \bibnamefont{Ensher}}, \bibinfo {author}
  {\bibfnamefont{C.~E.}\ \bibnamefont{Wieman}},\ and\ \bibinfo {author}
  {\bibfnamefont{E.~A.}\ \bibnamefont{Cornell}},\ }%
  \bibfield{journal}{%
  \bibinfo {journal} {Phys. Rev. Lett.}\ }%
  \textbf{\bibinfo {volume} {81}},\ \bibinfo {pages} {1539} (\bibinfo {year}
  {1998})%
  \bibAnnoteFile{NoStop}{hall98}%
\bibitem{sadler06}%
  \BibitemOpen
  \bibfield{author}{%
  \bibinfo {author} {\bibfnamefont{L.~E.}\ \bibnamefont{Sadler}}, \bibinfo
  {author} {\bibfnamefont{J.~M.}\ \bibnamefont{Higbie}}, \bibinfo {author}
  {\bibfnamefont{S.~R.}\ \bibnamefont{Leslie}}, \bibinfo {author}
  {\bibfnamefont{M.}~\bibnamefont{Vengalattore}},\ and\ \bibinfo {author}
  {\bibfnamefont{D.~M.}\ \bibnamefont{Stamper-Kurn}},\ }%
  \bibfield{journal}{%
  \bibinfo {journal} {Nature}\ }%
  \textbf{\bibinfo {volume} {443}},\ \bibinfo {pages} {312} (\bibinfo {year}
  {2006})%
  \bibAnnoteFile{NoStop}{sadler06}%
\bibitem{guzman11}%
  \BibitemOpen
  \bibfield{author}{%
  \bibinfo {author} {\bibfnamefont{J.}~\bibnamefont{Guzman}}, \bibinfo {author}
  {\bibfnamefont{G.-B.}\ \bibnamefont{Jo}}, \bibinfo {author}
  {\bibfnamefont{A.~N.}\ \bibnamefont{Wenz}}, \bibinfo {author}
  {\bibfnamefont{K.~W.}\ \bibnamefont{Murch}}, \bibinfo {author}
  {\bibfnamefont{C.~K.}\ \bibnamefont{Thomas}},\ and\ \bibinfo {author}
  {\bibfnamefont{D.~M.}\ \bibnamefont{Stamper-Kurn}},\ }%
  \bibfield{journal}{%
  \bibinfo {journal} {Phys. Rev. A}\ }%
  \textbf{\bibinfo {volume} {84}},\ \bibinfo {pages} {063625} (\bibinfo {year}
  {2011})%
  \bibAnnoteFile{NoStop}{guzman11}%
\bibitem{de13}%
  \BibitemOpen
  \bibfield{author}{%
  \bibinfo {author} {\bibfnamefont{S.}~\bibnamefont{De}}, \bibinfo {author}
  {\bibfnamefont{D.~L.}\ \bibnamefont{Campbell}}, \bibinfo {author}
  {\bibfnamefont{R.~M.}\ \bibnamefont{Price}}, \bibinfo {author}
  {\bibfnamefont{A.}~\bibnamefont{Putra}}, \bibinfo {author}
  {\bibfnamefont{B.~M.}\ \bibnamefont{Anderson}},\ and\ \bibinfo {author}
  {\bibfnamefont{I.~B.}\ \bibnamefont{Spielman}},\ }%
  \bibfield{journal}{%
  \bibinfo {journal} {Phys. Rev. A}\ }%
  \textbf{\bibinfo {volume} {89}},\ \bibinfo {pages} {033631} (\bibinfo {month}
  {Mar}\ \bibinfo {year} {2014})%
  \bibAnnoteFile{NoStop}{de13}%
\bibitem{papp08}%
  \BibitemOpen
  \bibfield{author}{%
  \bibinfo {author} {\bibfnamefont{S.~B.}\ \bibnamefont{Papp}}, \bibinfo
  {author} {\bibfnamefont{J.~M.}\ \bibnamefont{Pino}},\ and\ \bibinfo {author}
  {\bibfnamefont{C.~E.}\ \bibnamefont{Wieman}},\ }%
  \bibfield{journal}{%
  \bibinfo {journal} {Phys. Rev. Lett.}\ }%
  \textbf{\bibinfo {volume} {101}},\ \bibinfo {pages} {040402} (\bibinfo
  {month} {Jul}\ \bibinfo {year} {2008})%
  \bibAnnoteFile{NoStop}{papp08}%
\bibitem{stamper13}%
  \BibitemOpen
  \bibfield{author}{%
  \bibinfo {author} {\bibfnamefont{D.~M.}\ \bibnamefont{Stamper-Kurn}}\ and\
  \bibinfo {author} {\bibfnamefont{M.}~\bibnamefont{Ueda}},\ }%
  \bibfield{journal}{%
  \bibinfo {journal} {Rev. Mod. Phys.}\ }%
  \textbf{\bibinfo {volume} {85}},\ \bibinfo {pages} {1191} (\bibinfo {year}
  {2013})%
  \bibAnnoteFile{NoStop}{stamper13}%
\bibitem{graf67}%
  \BibitemOpen
  \bibfield{author}{%
  \bibinfo {author} {\bibfnamefont{E.~H.}\ \bibnamefont{Graf}}, \bibinfo
  {author} {\bibfnamefont{D.~M.}\ \bibnamefont{Lee}},\ and\ \bibinfo {author}
  {\bibfnamefont{J.~D.}\ \bibnamefont{Reppy}},\ }%
  \bibfield{journal}{%
  \bibinfo {journal} {Phys. Rev. Lett.}\ }%
  \textbf{\bibinfo {volume} {19}},\ \bibinfo {pages} {417} (\bibinfo {month}
  {Aug}\ \bibinfo {year} {1967})%
  \bibAnnoteFile{NoStop}{graf67}%
\bibitem{parker13}%
  \BibitemOpen
  \bibfield{author}{%
  \bibinfo {author} {\bibfnamefont{C.~V.}\ \bibnamefont{Parker}}, \bibinfo
  {author} {\bibfnamefont{L.-C.}\ \bibnamefont{Ha}},\ and\ \bibinfo {author}
  {\bibfnamefont{C.}~\bibnamefont{Chin}},\ }%
  \bibfield{journal}{%
  \bibinfo {journal} {Nature Physics}\ }%
  \textbf{\bibinfo {volume} {9}},\ \bibinfo {pages} {769} (\bibinfo {year}
  {2013})%
  \bibAnnoteFile{NoStop}{parker13}%
\bibitem{karl13}%
  \BibitemOpen
  \bibfield{author}{%
  \bibinfo {author} {\bibfnamefont{M.}~\bibnamefont{Karl}}, \bibinfo {author}
  {\bibfnamefont{B.}~\bibnamefont{Nowak}},\ and\ \bibinfo {author}
  {\bibfnamefont{T.}~\bibnamefont{Gasenzer}},\ }%
  \bibfield{journal}{%
  \bibinfo {journal} {Phys. Rev. A}\ }%
  \textbf{\bibinfo {volume} {88}},\ \bibinfo {pages} {063615} (\bibinfo {year}
  {2013})%
  \bibAnnoteFile{NoStop}{karl13}%
\bibitem{karl132}%
  \BibitemOpen
  \bibfield{author}{%
  \bibinfo {author} {\bibfnamefont{M.}~\bibnamefont{Karl}}, \bibinfo {author}
  {\bibfnamefont{B.}~\bibnamefont{Nowak}},\ and\ \bibinfo {author}
  {\bibfnamefont{T.}~\bibnamefont{Gasenzer}},\ }%
  \bibfield{journal}{%
  \bibinfo {journal} {Scientific Reports}\ }%
  \textbf{\bibinfo {volume} {3}},\ \bibinfo {pages} {2394} (\bibinfo {year}
  {2013})%
  \bibAnnoteFile{NoStop}{karl132}%
\bibitem{nowak12}%
  \BibitemOpen
  \bibfield{author}{%
  \bibinfo {author} {\bibfnamefont{B.}~\bibnamefont{Nowak}}, \bibinfo {author}
  {\bibfnamefont{J.}~\bibnamefont{Schole}}, \bibinfo {author}
  {\bibfnamefont{D.}~\bibnamefont{Sexty}},\ and\ \bibinfo {author}
  {\bibfnamefont{T.}~\bibnamefont{Gasenzer}},\ }%
  \bibfield{journal}{%
  \bibinfo {journal} {Phys. Rev. A}\ }%
  \textbf{\bibinfo {volume} {85}},\ \bibinfo {pages} {043627} (\bibinfo {year}
  {2012})%
  \bibAnnoteFile{NoStop}{nowak12}%
\bibitem{nowak11}%
  \BibitemOpen
  \bibfield{author}{%
  \bibinfo {author} {\bibfnamefont{B.}~\bibnamefont{Nowak}}, \bibinfo {author}
  {\bibfnamefont{D.}~\bibnamefont{Sexty}},\ and\ \bibinfo {author}
  {\bibfnamefont{T.}~\bibnamefont{Gasenzer}},\ }%
  \bibfield{journal}{%
  \bibinfo {journal} {Phys. Rev. B (R)}\ }%
  \textbf{\bibinfo {volume} {84}},\ \bibinfo {pages} {020506} (\bibinfo {year}
  {2011})%
  \bibAnnoteFile{NoStop}{nowak11}%
\bibitem{schole12}%
  \BibitemOpen
  \bibfield{author}{%
  \bibinfo {author} {\bibfnamefont{J.}~\bibnamefont{Schole}}, \bibinfo {author}
  {\bibfnamefont{B.}~\bibnamefont{Nowak}},\ and\ \bibinfo {author}
  {\bibfnamefont{T.}~\bibnamefont{Gasenzer}},\ }%
  \bibfield{journal}{%
  \bibinfo {journal} {Phys. Rev. A}\ }%
  \textbf{\bibinfo {volume} {86}},\ \bibinfo {pages} {013624} (\bibinfo {year}
  {2012})%
  \bibAnnoteFile{NoStop}{schole12}%
\bibitem{kasamatsu04}%
  \BibitemOpen
  \bibfield{author}{%
  \bibinfo {author} {\bibfnamefont{K.}~\bibnamefont{Kasamatsu}}\ and\ \bibinfo
  {author} {\bibfnamefont{M.}~\bibnamefont{Tsubota}},\ }%
  \bibfield{journal}{%
  \bibinfo {journal} {Phys. Rev. Lett.}\ }%
  \textbf{\bibinfo {volume} {93}},\ \bibinfo {pages} {100402} (\bibinfo {year}
  {2004})%
  \bibAnnoteFile{NoStop}{kasamatsu04}%
\bibitem{lamacraft07}%
  \BibitemOpen
  \bibfield{author}{%
  \bibinfo {author} {\bibfnamefont{A.}~\bibnamefont{Lamacraft}},\ }%
  \bibfield{journal}{%
  \bibinfo {journal} {Phys. Rev. Lett.}\ }%
  \textbf{\bibinfo {volume} {98}},\ \bibinfo {pages} {160404} (\bibinfo {month}
  {Apr}\ \bibinfo {year} {2007})%
  \bibAnnoteFile{NoStop}{lamacraft07}%
\bibitem{ronen08}%
  \BibitemOpen
  \bibfield{author}{%
  \bibinfo {author} {\bibfnamefont{S.}~\bibnamefont{Ronen}}, \bibinfo {author}
  {\bibfnamefont{J.~L.}\ \bibnamefont{Bohn}}, \bibinfo {author}
  {\bibfnamefont{L.~E.}\ \bibnamefont{Halmo}},\ and\ \bibinfo {author}
  {\bibfnamefont{M.}~\bibnamefont{Edwards}},\ }%
  \bibfield{journal}{%
  \bibinfo {journal} {Phys. Rev. A}\ }%
  \textbf{\bibinfo {volume} {78}},\ \bibinfo {pages} {053613} (\bibinfo {month}
  {Nov}\ \bibinfo {year} {2008})%
  \bibAnnoteFile{NoStop}{ronen08}%
\bibitem{barnett11}%
  \BibitemOpen
  \bibfield{author}{%
  \bibinfo {author} {\bibfnamefont{R.}~\bibnamefont{Barnett}}, \bibinfo
  {author} {\bibfnamefont{A.}~\bibnamefont{Polkovnikov}},\ and\ \bibinfo
  {author} {\bibfnamefont{M.}~\bibnamefont{Vengalattore}},\ }%
  \bibfield{journal}{%
  \bibinfo {journal} {Phys. Rev. A}\ }%
  \textbf{\bibinfo {volume} {84}},\ \bibinfo {pages} {023606} (\bibinfo {month}
  {Aug}\ \bibinfo {year} {2011})%
  \bibAnnoteFile{NoStop}{barnett11}%
\bibitem{mukerjee07}%
  \BibitemOpen
  \bibfield{author}{%
  \bibinfo {author} {\bibfnamefont{S.}~\bibnamefont{Mukerjee}}, \bibinfo
  {author} {\bibfnamefont{C.}~\bibnamefont{Xu}},\ and\ \bibinfo {author}
  {\bibfnamefont{J.~E.}\ \bibnamefont{Moore}},\ }%
  \bibfield{journal}{%
  \bibinfo {journal} {Phys. Rev. B}\ }%
  \textbf{\bibinfo {volume} {76}},\ \bibinfo {pages} {104519} (\bibinfo {month}
  {Sep}\ \bibinfo {year} {2007})%
  \bibAnnoteFile{NoStop}{mukerjee07}%
\bibitem{kudo13}%
  \BibitemOpen
  \bibfield{author}{%
  \bibinfo {author} {\bibfnamefont{K.}~\bibnamefont{Kudo}}\ and\ \bibinfo
  {author} {\bibfnamefont{Y.}~\bibnamefont{Kawaguchi}},\ }%
  \bibfield{journal}{%
  \bibinfo {journal} {Phys. Rev. A}\ }%
  \textbf{\bibinfo {volume} {88}},\ \bibinfo {pages} {013630} (\bibinfo {month}
  {Jul}\ \bibinfo {year} {2013})%
  \bibAnnoteFile{NoStop}{kudo13}%
\bibitem{damle96}%
  \BibitemOpen
  \bibfield{author}{%
  \bibinfo {author} {\bibfnamefont{K.}~\bibnamefont{Damle}}, \bibinfo {author}
  {\bibfnamefont{S.~N.}\ \bibnamefont{Majumdar}},\ and\ \bibinfo {author}
  {\bibfnamefont{S.}~\bibnamefont{Sachdev}},\ }%
  \bibfield{journal}{%
  \bibinfo {journal} {Phys. Rev. A}\ }%
  \textbf{\bibinfo {volume} {54}},\ \bibinfo {pages} {5037} (\bibinfo {month}
  {Dec}\ \bibinfo {year} {1996})%
  \bibAnnoteFile{NoStop}{damle96}%
\bibitem{kempen02}%
  \BibitemOpen
  \bibfield{author}{%
  \bibinfo {author} {\bibfnamefont{E.~G.~M.}\ \bibnamefont{van Kempen}},
  \bibinfo {author} {\bibfnamefont{S.~J. J. M.~F.}\ \bibnamefont{Kokkelmans}},
  \bibinfo {author} {\bibfnamefont{D.~J.}\ \bibnamefont{Heinzen}},\ and\
  \bibinfo {author} {\bibfnamefont{B.~J.}\ \bibnamefont{Verhaar}},\ }%
  \bibfield{journal}{%
  \bibinfo {journal} {Phys. Rev. Lett.}\ }%
  \textbf{\bibinfo {volume} {88}},\ \bibinfo {pages} {093201} (\bibinfo {year}
  {2002})%
  \bibAnnoteFile{NoStop}{kempen02}%
\bibitem{colson78}%
  \BibitemOpen
  \bibfield{author}{%
  \bibinfo {author} {\bibfnamefont{W.~B.}\ \bibnamefont{Colson}}\ and\ \bibinfo
  {author} {\bibfnamefont{A.~L.}\ \bibnamefont{Fetter}},\ }%
  \bibfield{journal}{%
  \bibinfo {journal} {Journal of Low Temperature Physics}\ }%
  \textbf{\bibinfo {volume} {33}},\ \bibinfo {pages} {231} (\bibinfo {year}
  {1978})%
  \bibAnnoteFile{NoStop}{colson78}%
\bibitem{law97}%
  \BibitemOpen
  \bibfield{author}{%
  \bibinfo {author} {\bibfnamefont{C.~K.}\ \bibnamefont{Law}}, \bibinfo
  {author} {\bibfnamefont{H.}~\bibnamefont{Pu}}, \bibinfo {author}
  {\bibfnamefont{N.~P.}\ \bibnamefont{Bigelow}},\ and\ \bibinfo {author}
  {\bibfnamefont{J.~H.}\ \bibnamefont{Eberly}},\ }%
  \bibfield{journal}{%
  \bibinfo {journal} {Phys. Rev. Lett.}\ }%
  \textbf{\bibinfo {volume} {79}},\ \bibinfo {pages} {3105} (\bibinfo {month}
  {Oct}\ \bibinfo {year} {1997})%
  \bibAnnoteFile{NoStop}{law97}%
\bibitem{stamper99}%
  \BibitemOpen
  \bibfield{author}{%
  \bibinfo {author} {\bibfnamefont{D.~M.}\ \bibnamefont{Stamper-Kurn}},
  \bibinfo {author} {\bibfnamefont{H.-J.}\ \bibnamefont{Miesner}}, \bibinfo
  {author} {\bibfnamefont{A.~P.}\ \bibnamefont{Chikkatur}}, \bibinfo {author}
  {\bibfnamefont{S.}~\bibnamefont{Inouye}}, \bibinfo {author}
  {\bibfnamefont{J.}~\bibnamefont{Stenger}},\ and\ \bibinfo {author}
  {\bibfnamefont{W.}~\bibnamefont{Ketterle}},\ }%
  \bibfield{journal}{%
  \bibinfo {journal} {Phys. Rev. Lett.}\ }%
  \textbf{\bibinfo {volume} {83}},\ \bibinfo {pages} {661} (\bibinfo {year}
  {1999})%
  \bibAnnoteFile{NoStop}{stamper99}%
\bibitem{chengreview2010}%
  \BibitemOpen
  \bibfield{author}{%
  \bibinfo {author} {\bibfnamefont{C.}~\bibnamefont{Chin}}, \bibinfo {author}
  {\bibfnamefont{R.}~\bibnamefont{Grimm}}, \bibinfo {author}
  {\bibfnamefont{P.}~\bibnamefont{Julienne}},\ and\ \bibinfo {author}
  {\bibfnamefont{E.}~\bibnamefont{Tiesinga}},\ }%
  \bibfield{journal}{%
  \bibinfo {journal} {Rev. Mod. Phys.}\ }%
  \textbf{\bibinfo {volume} {82}},\ \bibinfo {pages} {1225} (\bibinfo {month}
  {Apr}\ \bibinfo {year} {2010})%
  \bibAnnoteFile{NoStop}{chengreview2010}%
\bibitem{bao03}%
  \BibitemOpen
  \bibfield{author}{%
  \bibinfo {author} {\bibfnamefont{W.}~\bibnamefont{Bao}}, \bibinfo {author}
  {\bibfnamefont{D.}~\bibnamefont{Jaksch}},\ and\ \bibinfo {author}
  {\bibfnamefont{P.~A.}\ \bibnamefont{Markowich}},\ }%
  \bibfield{journal}{%
  \bibinfo {journal} {Journal of Computational Physics}\ }%
  \textbf{\bibinfo {volume} {187}},\ \bibinfo {pages} {318} (\bibinfo {year}
  {2003})%
  \bibAnnoteFile{NoStop}{bao03}%
\bibitem{weber02}%
  \BibitemOpen
  \bibfield{author}{%
  \bibinfo {author} {\bibfnamefont{T.}~\bibnamefont{Weber}}, \bibinfo {author}
  {\bibfnamefont{J.}~\bibnamefont{Herbig}}, \bibinfo {author}
  {\bibfnamefont{M.}~\bibnamefont{Mark}}, \bibinfo {author}
  {\bibfnamefont{H.-C.}\ \bibnamefont{Nagerl}},\ and\ \bibinfo {author}
  {\bibfnamefont{G.}~\bibnamefont{R.}},\ }%
  \bibfield{journal}{%
  \bibinfo {journal} {Science}\ }%
  \textbf{\bibinfo {volume} {299}},\ \bibinfo {pages} {232} (\bibinfo {year}
  {2002})%
  \bibAnnoteFile{NoStop}{weber02}%
\bibitem{vuleti99}%
  \BibitemOpen
  \bibfield{author}{%
  \bibinfo {author}
  {\bibfnamefont{V.}~\bibnamefont{Vuleti\ifmmode~\acute{c}\else \'{c}\fi{}}},
  \bibinfo {author} {\bibfnamefont{A.~J.}\ \bibnamefont{Kerman}}, \bibinfo
  {author} {\bibfnamefont{C.}~\bibnamefont{Chin}},\ and\ \bibinfo {author}
  {\bibfnamefont{S.}~\bibnamefont{Chu}},\ }%
  \bibfield{journal}{%
  \bibinfo {journal} {Phys. Rev. Lett.}\ }%
  \textbf{\bibinfo {volume} {82}},\ \bibinfo {pages} {1406} (\bibinfo {month}
  {Feb}\ \bibinfo {year} {1999})%
  \bibAnnoteFile{NoStop}{vuleti99}%
\bibitem{pollack09}%
  \BibitemOpen
  \bibfield{author}{%
  \bibinfo {author} {\bibfnamefont{S.~E.}\ \bibnamefont{Pollack}}, \bibinfo
  {author} {\bibfnamefont{D.}~\bibnamefont{Dries}}, \bibinfo {author}
  {\bibfnamefont{M.}~\bibnamefont{Junker}}, \bibinfo {author}
  {\bibfnamefont{Y.~P.}\ \bibnamefont{Chen}}, \bibinfo {author}
  {\bibfnamefont{T.~A.}\ \bibnamefont{Corcovilos}},\ and\ \bibinfo {author}
  {\bibfnamefont{R.~G.}\ \bibnamefont{Hulet}},\ }%
  \bibfield{journal}{%
  \bibinfo {journal} {Phys. Rev. Lett.}\ }%
  \textbf{\bibinfo {volume} {102}},\ \bibinfo {pages} {090402} (\bibinfo
  {month} {Mar}\ \bibinfo {year} {2009})%
  \bibAnnoteFile{NoStop}{pollack09}%
\bibitem{huse86}%
  \BibitemOpen
  \bibfield{author}{%
  \bibinfo {author} {\bibfnamefont{D.~A.}\ \bibnamefont{Huse}},\ }%
  \bibfield{journal}{%
  \bibinfo {journal} {Phys. Rev. B}\ }%
  \textbf{\bibinfo {volume} {34}},\ \bibinfo {pages} {7845} (\bibinfo {year}
  {1986})%
  \bibAnnoteFile{NoStop}{huse86}%
\bibitem{furukawa85}%
  \BibitemOpen
  \bibfield{author}{%
  \bibinfo {author} {\bibfnamefont{H.}~\bibnamefont{Furukawa}},\ }%
  \bibfield{journal}{%
  \bibinfo {journal} {Phys. Rev. A}\ }%
  \textbf{\bibinfo {volume} {31}},\ \bibinfo {pages} {1103} (\bibinfo {year}
  {1985})%
  \bibAnnoteFile{NoStop}{furukawa85}%
\bibitem{bert08}%
  \BibitemOpen
  \bibfield{author}{%
  \bibinfo {author} {\bibfnamefont{B.}~\bibnamefont{Van~Schaeybroeck}},\ }%
  \bibfield{journal}{%
  \bibinfo {journal} {Phys. Rev. A}\ }%
  \textbf{\bibinfo {volume} {78}},\ \bibinfo {pages} {023624} (\bibinfo {month}
  {Aug}\ \bibinfo {year} {2008})%
  \bibAnnoteFile{NoStop}{bert08}%
\bibitem{siggia79}%
  \BibitemOpen
  \bibfield{author}{%
  \bibinfo {author} {\bibfnamefont{E.~D.}\ \bibnamefont{Siggia}},\ }%
  \bibfield{journal}{%
  \bibinfo {journal} {Phys. Rev. A}\ }%
  \textbf{\bibinfo {volume} {20}},\ \bibinfo {pages} {595} (\bibinfo {month}
  {Aug}\ \bibinfo {year} {1979})%
  \bibAnnoteFile{NoStop}{siggia79}%
\bibitem{wagner98}%
  \BibitemOpen
  \bibfield{author}{%
  \bibinfo {author} {\bibfnamefont{A.~J.}\ \bibnamefont{Wagner}}\ and\ \bibinfo
  {author} {\bibfnamefont{J.~M.}\ \bibnamefont{Yeomans}},\ }%
  \bibfield{journal}{%
  \bibinfo {journal} {Phys. Rev. Lett.}\ }%
  \textbf{\bibinfo {volume} {80}},\ \bibinfo {pages} {1429} (\bibinfo {month}
  {Feb}\ \bibinfo {year} {1998})%
  \bibAnnoteFile{NoStop}{wagner98}%
\bibitem{berloff02}%
  \BibitemOpen
  \bibfield{author}{%
  \bibinfo {author} {\bibfnamefont{N.~G.}\ \bibnamefont{Berloff}}\ and\
  \bibinfo {author} {\bibfnamefont{B.~V.}\ \bibnamefont{Svistunov}},\ }%
  \bibfield{journal}{%
  \bibinfo {journal} {Phys. Rev. A}\ }%
  \textbf{\bibinfo {volume} {66}},\ \bibinfo {pages} {013603} (\bibinfo {month}
  {Jul}\ \bibinfo {year} {2002})%
  \bibAnnoteFile{NoStop}{berloff02}%
\bibitem{blakie08}%
  \BibitemOpen
  \bibfield{author}{%
  \bibinfo {author} {\bibfnamefont{P.}~\bibnamefont{Blakie}}, \bibinfo {author}
  {\bibfnamefont{A.~S.}\ \bibnamefont{Bradley}}, \bibinfo {author}
  {\bibfnamefont{M.~J.}\ \bibnamefont{Davis}}, \bibinfo {author}
  {\bibfnamefont{R.~J.}\ \bibnamefont{Ballagh}},\ and\ \bibinfo {author}
  {\bibfnamefont{C.~W.}\ \bibnamefont{Gardiner}},\ }%
  \bibfield{journal}{%
  \bibinfo {journal} {Advances in Physics}\ }%
  \textbf{\bibinfo {volume} {57}},\ \bibinfo {pages} {363} (\bibinfo {year}
  {2008})%
  \bibAnnoteFile{NoStop}{blakie08}%
\bibitem{miyake11}%
  \BibitemOpen
  \bibfield{author}{%
  \bibinfo {author} {\bibfnamefont{H.}~\bibnamefont{Miyake}}, \bibinfo {author}
  {\bibfnamefont{G.}~\bibnamefont{Siviloglou}}, \bibinfo {author}
  {\bibfnamefont{G.}~\bibnamefont{Puentes}}, \bibinfo {author}
  {\bibfnamefont{D.~E.}\ \bibnamefont{Pritchard}}, \bibinfo {author}
  {\bibfnamefont{W.}~\bibnamefont{Ketterle}},\ and\ \bibinfo {author}
  {\bibfnamefont{D.~M.}\ \bibnamefont{Weld}},\ }%
  \bibfield{journal}{%
  \bibinfo {journal} {Phys. Rev. Lett.}\ }%
  \textbf{\bibinfo {volume} {107}},\ \bibinfo {pages} {175302} (\bibinfo {year}
  {2011})%
  \bibAnnoteFile{NoStop}{miyake11}%
\end{thebibliography}%

\end{document}